\documentstyle[aps,prl,floats,epsf,twocolumn]{revtex}

\begin{document}

\wideabs{

\title{Observation of a Linearly Dispersing Collective Mode in a Quantum Hall 
Ferromagnet}

\author{I.B.~Spielman$^1$, J.~P. Eisenstein$^1$, 
L.~N. Pfeiffer$^2$, and K. W. West$^2$}

\address{$^1$California Institute of Technology, Pasadena CA 91125 \\
         $^2$Bell Laboratories, Lucent Technologies, Murray Hill, NJ 07974}

\maketitle

\begin{abstract}

Double layer two-dimensional electron systems can exhibit a fascinating 
collective phase believed to exhibit both quantum ferromagnetism and excitonic 
superfluidity.  This unusual  phase has recently been found to exhibit
tunneling phenomena reminiscent of the  Josephson effect.  A key element of the
theoretical understanding of this  bizarre quantum fluid is the existence of
linearly dispersing Goldstone  collective modes. Using the method of tunneling
spectroscopy, we have  demonstrated the existence of these modes. We find the
measured velocity to be  in reasonable agreement with theoretical estimates.

\end{abstract}

} 

The remarkable properties of superfluids and superconductors are intimately 
related to existence of a bosonic condensate of composite particles consisting 
of an even number of fermions within the strongly interacting many-body 
environment.  In superfluid $\rm{^4He}$ these composite particles are the
helium atoms themselves.  In superconductors Cooper pairs play the analogous
role. In  semiconductors excitons, which consist of conduction band electrons
paired with  valence band holes, have long been considered candidate building
blocks of a new  class of neutral superfluids\cite{keldysh}.  More recently, it
has become apparent that double  layer two-dimensional electron systems provide
yet another system for realizing  the exotica of superfluidity.  In the
presence of a large magnetic field $B$  such systems will exhibit a quantized
Hall plateau at $\rho_{xy}$ = $h/e^2$ if  the layer separation is sufficiently
small\cite{perspectives}.   When tunneling  between the layers is weak this
quantum Hall state reflects the condensation of a remarkable strongly
interacting electron quantum fluid.  This fluid may be  viewed as an excitonic
superfluid in which an electron in one layer is paired with a hole (in
the {\it conduction} band) in the other layer.   Quantum mechanical uncertainty
makes it impossible to tell which layer either  component of this composite
boson is in.  Equivalently, the system may be regarded as a ferromagnet in
which every electron exists in a coherent  superposition of the ``pseudospin''
eigenstates which encode the layer degree of  freedom\cite{yang}.  The phase
variable of this superposition, which is  analogous to the phase variable in
conventional superconductors or superfluid  $\rm{^4He}$, determines the
orientation of the pseudospin magnetic moment.  Spatial variations of the phase
govern the low energy excitations in the system.   We report here strong
evidence, obtained via tunneling spectroscopy, for these  excitations.

The physical properties of this bilayer quantum fluid are predicted to be quite 
exotic\cite{perspectives}.  For example, the transition to the ordered state is
expected to be a finite temperature Kosterlitz-Thouless (KT) transition.  This is
unusual since the quantum Hall effect (QHE) is usually only associated with a
zero temperature quantum phase transition. Below the KT transition temperature, 
$T_{KT}$, counterflow superfluidity is expected:  transport involving equal but 
oppositely directed currents in the two layers should be dissipationless, at 
least in linear response\cite{perspectives,stern0}.  There have also been 
controversial predictions of a Josephson effect in the tunneling transport
between the layers\cite{wenzee,ezawa}.  This issue in particular has come  under
renewed scrutiny since the experimental discovery of a huge resonant  enhancement
of the zero bias tunneling conductance when the separation between  the layers is
reduced below a critical value\cite{spielman}.

The Hall plateau $\rho_{xy}$ = $h/e^2$ corresponds to total Landau level
filling  factor $\nu_T=1$, with $\nu_T$ defined as the ratio of the total
electron  density $N_T$ to the degeneracy $eB/h$ of a single spin resolved
Landau level.  If the layers are identical, this occurs despite the fact that
each layer viewed  independently has filling factor $\nu=1/2$. That the net
system exhibits a QHE  even though the layers by themselves do not, is a result
of the interlayer  couplings in the system.  If tunneling is negligible the
$\nu_T=1$ QHE results  from the interplay of Coulomb interactions among
electrons in the same layer  with those in opposite layers. If the separation
between the layers is too  large, the interlayer correlations break down and
the QHE disappears. 

A good approximation to the ground state at small layer separation is the 
product of a Slater determinant of all orbital states in the lowest Landau 
level and a totally symmetric pseudospin wavefunction\cite{true spins}.  An 
electron in one layer is pseudospin up, $|\!\!\uparrow \rangle$, while one in 
the other layer is pseudospin down, $|\!\!\downarrow \rangle$.  Exchange 
interactions favor each electron being in the same pseudospin state:  
$|\!\uparrow \rangle + e^{i \phi} |\!\downarrow \rangle$.  The phase $\phi$ is 
uniform and, in the absence of tunneling, arbitrary.  The system is thus  an
easy-plane ferromagnet whose moment lies near the $xy$-plane of pseudospin 
space. Owing to the finite layer separation the pseudospin moment fluctuates, 
both within the plane and perpendicular to it.  The latter corresponds to
local  fluctuations in the density difference between the layers and are
attended by a  capacitive energy penalty.  These fluctuations become
increasingly severe at  larger layer spacing and eventually destroy the ordered
state.  The precise  nature of the associated quantum phase transition is still
being vigorously  investigated\cite{schliemann,demler,kim}. Although the
ferromagnetic phase has a gap to charged excitations (and hence displays a
QHE), it also possesses  linearly dispersing neutral collective modes (i.e.
pseudospin waves) associated  with spatial gradients in the phase. These are
the Goldstone modes of the broken  symmetry ground state and are gapless in the
long wavelength  limit\cite{fertig,wenzee}.  These modes are the main focus of
this paper.

\begin{figure}
\begin{center}
\epsfclipon
\epsffile{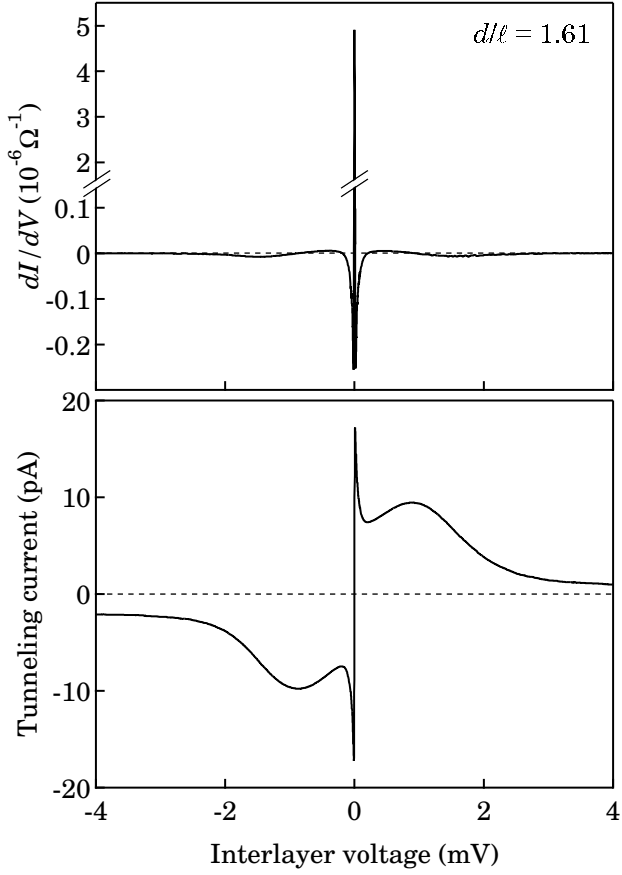}
\end{center}
\caption[figure 1]{
Tunneling data at $\nu_T=1$ and $T=25$mK.  Total density $N_T = 5.2\times 10^{10}cm^{-2}$.  Upper panel: Conductance $dI/dV$ {\it vs.} 
interlayer voltage $V$; lower panel: Tunnel current $I$ {\it vs.} $V$.}
\label{Zero}
\end{figure}

Figure \ref{Zero} shows the measured tunneling conductance $dI/dV$ and current 
$I$ {\it vs.} the voltage difference $V$ between two 2D electron gas layers at
$\nu_T=1$ at  $T=25$mK. The sample consists of two individually contacted
modulation-doped 18nm GaAs quantum wells separated by a 9.9nm $\rm
{Al_{0.9}Ga_{0.1}As}$ barrier layer.  Electrostatic gating is used to adjust the
layer densities; for the data shown $N_1\!=\!N_2\!=\!N_T/2\!=\!2.6 \times
10^{10} cm^{-2}$.  At this density the ratio  of interlayer separation $d$ to the
magnetic length $\ell =(\hbar /eB)^{1/2}$ is $d/\ell =1.61$.  This ratio
determines the relative importance of inter- and intra-layer Coulomb interactions
in the system. As Spielman, {\it et  al.}\cite{spielman} showed, the sharp peak
in $dI/dV$ at $V=0$ disappears when the density, and thus $d/\ell$, is increased.
Above $d/\ell  \approx 1.84$ the peak in $dI/dV$ is replaced by the familiar
deep  minimum characteristic of the Coulomb barrier to tunneling between {\it 
uncorrelated} 2D systems\cite{jpe1}.

\begin{figure}
\begin{center}
\epsfclipon
\epsffile{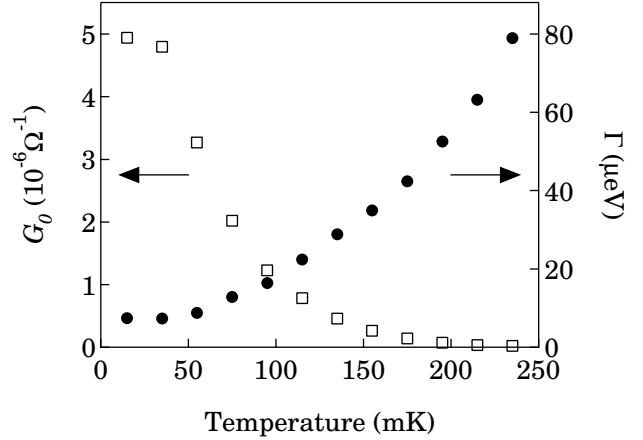}
\end{center}
\caption[figure 2]{
Linewidth $\Gamma$ (solid dots) and peak height $G_0$ (open squares) at
$\nu_T=1$.  Total density $N_T = 5.2\times 10^{10}cm^{-2}$.
}
\label{TempDependance}
\end{figure}

The sharp zero bias peak in $dI/dV$ illustrated in Fig. \ref{Zero} is accompanied
by a near discontinuity in the tunnel current $I$.  Although this observation is
suggestive of a Josephson effect, the present data do not exhibit a true
supercurrent at zero bias.  The zero bias conductance $G_0$, while  vastly
enhanced over its value at higher densities, remains finite as 
$T\!\!\rightarrow\!0$.  Similarly, the width $\Gamma$ of the zero bias
conductance peak attains a minimum, but non-zero, value in this limit. 

Figure \ref{TempDependance} illustrates the temperature dependence of the zero
bias conductance $G_0$ and peak width $\Gamma$ at $d/\ell=1.61$.  The peak
height rises steadily as the temperature is reduced to around 50mK.  Below this
temperature it saturates at $G_0 \approx 5 \times 10^{-6} \Omega^{-1}$.  The
width $\Gamma$ (defined as the full width at half maximum of the $dI/dV$ peak)
decreases down to about the same temperature below which it saturates at
$\Gamma \approx 6 \mu V$.  We emphasize that this resonance is roughly 15 times
narrower and 150 times taller than the tunnel resonance observed at zero
magnetic field in the same sample.  At zero field tunneling can be understood
in single particle terms and the linewidth reflects the lifetime of the
quasiparticles in the 2D systems\cite{murphy}. In contrast, the dramatic
resonance at $\nu_T=1$ suggests that a collective mode dominates the spectral
weight at low energy.

The low temperature saturation of $G_0$ and $\Gamma$ is not understood.  Indeed, it is not at all clear that the saturation is intrinsic. Extrinsic electromagnetic interference led to the broader and weaker zero bias peaks reported in the original work of Spielman, {\it et al.}\cite{spielman}. Although much effort has since been expended in improving the noise environment, the possibility of a still narrower conductance peak at very low temperature remains.  Electron heating is another potential source of the saturation. Finally, we remark that as the present tunneling measurements are effectively two-terminal, finite series resistances will ultimately limit height of the tunnel peak.   

\begin{figure}
\begin{center}
\epsfclipon
\epsffile{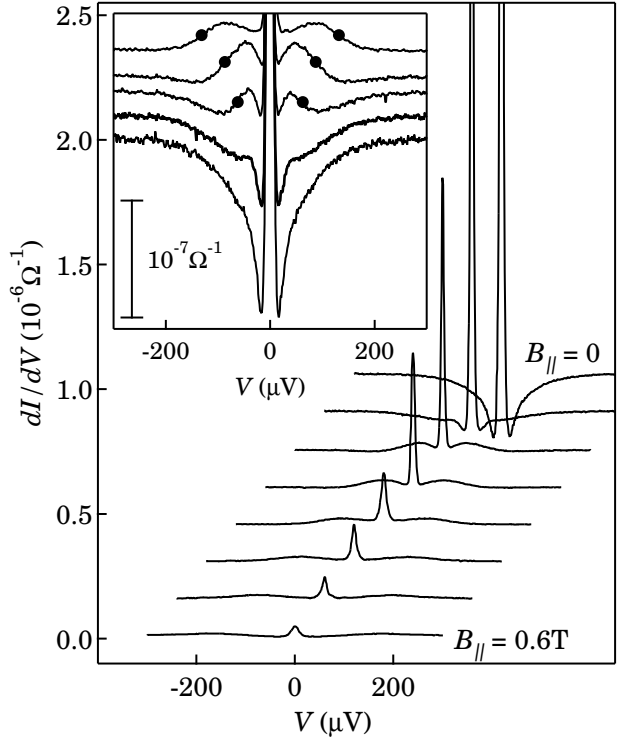}
\end{center}
\caption[figure 3]{

Tunneling conductance spectra at $T=25$mK and $N_T=5.2\times  10^{10}cm^{-2}$
for various parallel magnetic fields.  Main panel: $B_{||}=0$,  0.11, 0.24,
0.29, 0.35, 0.43, 0.49, and 0.59T.  Inset: Expanded view of  spectra for
$B_{||}=0.07$, 0.11, 0.15, 0.24, and 0.35T.  Dots indicate the positions of the
split-off resonances in $dI/dV$.

}
\label{Angles}
\end{figure}

The Goldstone mode of the coherent QHE ground state offers a natural way to 
understand the zero bias peak at $\nu_T=1$. This mode involves oscillations of 
the pseudospin magnetization in the $xy$-plane and along the $z$ axis of 
pseudospin space. In the absence of tunneling, the mode has zero energy at
zero  wavevector $q$.  Any tunneling, however, opens a small gap $\Delta_0$ at
$q=0$  and allows the mode to effectively transfer charge between the layers.
If  $\Delta_0$ is small enough, a zero bias enhancement of the tunneling
conductance is expected.  Within this framework, the sharp zero bias  peak in $dI/dV$ represents a direct spectroscopic detection of the Goldstone mode of the $\nu_T=1$ broken symmetry ground state. The significance of the height and width of this peak is less clear\cite{stern,radzihovsky}.

To test this interpretation, we have examined the tunneling spectra after adding
a small magnetic field component $B_{||}$ parallel to the 2D planes to the 
existing perpendicular field $B_{\perp}$. The in-plane field renders the 
$I\!\!-\!\!V$ characteristic sensitive to spectral features at the non-zero, and
{\it adjustable}, wavevector $q=eB_{||}d/\hbar$.  This powerful technique has
been applied widely in the past\cite{scalapino} and is of known efficacy in
tunneling between 2D electron systems in semiconductor 
heterostructures\cite{gornik,jpe2}.  In the present circumstance the parallel 
field has been predicted\cite{stern,radzihovsky} to split the zero bias 
tunneling peak into two resonances symmetric about $V=0$.  The voltage location 
of these resonances should be $eV=\pm\hbar\omega(q)$, where $\hbar\omega(q)$  is
the Goldstone mode energy at the parallel field-induced wavevector $q$. Since 
the mode disperses linearly for small $q$, detection of the splitting will 
provide a measure of its velocity $c$.  

Figure \ref{Angles} shows a sequence of tunnel spectra at $\nu_T=1$ and $T=25$mK
with  different parallel fields applied.  These data are obtained by tilting the
sample relative to an external magnetic field whose magnitude is adjusted to
maintain $B_{\perp}$, and hence the Landau level filling factor 
$\nu_T=hN_T/eB_{\perp}$ and magnetic length $\ell=(\hbar/eB_{\perp})^{1/2}$, 
constant.  The parallel field is accurately determined using a second 2D 
electron gas sample mounted perpendicular to the tunneling sample.  The Hall 
resistance of this second sample is then proportional to $B_{||}$.  The data 
shown in the Fig. \ref{Angles} again correspond to $N_1\!=\!N_2\!=\!N_T/2\!=\!5.2
\times  10^{10} cm^{-2}$ and $d/\ell=1.61$; similar data have been obtained at 
various other densities provided $d/\ell <1.84$.

It is clear from Fig. \ref{Angles} that the parallel magnetic field has a dramatic effect on the  tunnel spectrum.  Only a few tenths of a Tesla are required to strongly suppress the zero bias conductance peak.  Closer inspection, however, reveals a more subtle effect: complex structure appears on the flanks of the zero bias peak.  This structure, which is magnified in the inset to Fig. \ref{Angles}, first appears as two small peaks in $dI/dV$ positioned symmetrically about $V=0$, superimposed on the still substantial flanks of the main zero bias resonance.  As the parallel field increases these split-off peaks move toward higher energies and become more prominent.  At the same time the zero bias resonance weakens steadily.  As Fig. \ref{Angles} shows, the split-off peaks are really part of a more complex undulation in the conductance.  Qualitatively, these split-off resonances have the `derivative' shape theoretically expected\cite{stern,radzihovsky}. With this in mind, we identify the energy of the resonances with the voltage $V^*$ at which the derivative of the conductance, i.e. $d^2I/dV^2$, exhibits an extremum.  The solid dots in the inset to Fig. \ref{Angles} show this identification.  At high $B_{||}$ these intriguing resonances are lost in the broad, presumably incoherent, tunneling background.

\begin{figure}
\begin{center}
\epsfclipon
\epsffile{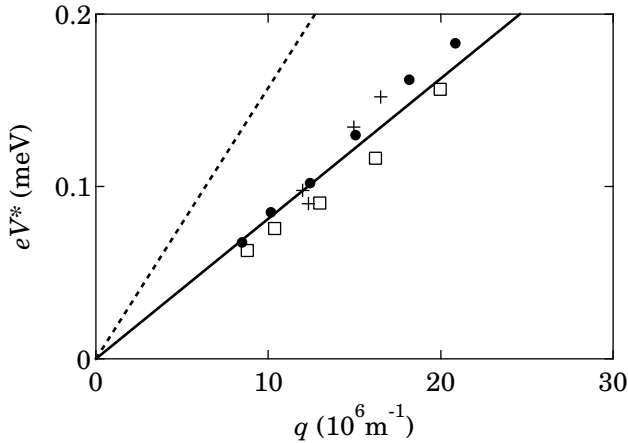}
\end{center}
\caption[figure 4]{

Energy, $eV^*$, of split-off peaks {\it vs.} the wavevector  $q=eB_{||}d/\hbar$
induced by the parallel magnetic field. Crosses, $N_T=6.4\times10^{10}cm^{-2}$;
empty squares,  $N_T=6.0\times10^{10}cm^{-2}$; filled circles,
$N_T=5.2\times10^{10}cm^{-2}$. Dashed line is a theoretical estimate\cite{macd} 
for the Goldstone mode dispersion relation at small $q$.  Solid line is a guide
to the eye and corresponds to a collective mode velocity of $1.4\times 10^4m/s$.

}
\label{Dispersion}
\end{figure}  

Figure \ref{Dispersion} displays the average energy $eV^*$ of the split-off resonances versus the wavevector $q=eB_{||}d/\hbar$.  Data for three different
densities,  $N_T=5.2$, 6.0, and $6.4\times 10^{10} cm^{-2}$, corresponding to
$d/\ell=1.61$,  1.71, and 1.76, are shown. Within the uncertainties these data
lie on straight lines whose slopes imply a velocity $c$ of about $1.4\times 10^4m/s$. We believe that these experimental results demonstrate the existence of a linearly dispersing collective mode in the bilayer 2D electron system at $\nu_T=1$.  This mode is very likely the anticipated pseudospin Goldstone mode of the broken symmetry state.  The dashed line in Fig. \ref{Dispersion} shows a recent theoretical estimate of the dispersion relation of this mode at long wavelengths\cite{macd}.    

The data in Figs. \ref{Angles} also possess aspects which are not explained by
the theoretical models\cite{stern,radzihovsky} mentioned above.  The biggest
puzzle is presented by the residual zero bias conductance peak which persists
to significant $B_{||}$. This feature is not contained within present
perturbative theories\cite{stern,radzihovsky}. One  possibility is that this
peak reflects second order (in the tunneling amplitude) ``two-magnon''
processes\cite{girvin}. Alternatively, the residual zero bias peak might be a
non-perturbative effect and one may speculate on its possible relation to a
Josephson supercurrent.  

An interesting analogy has been drawn\cite{fogler} between tunneling in this 
bilayer 2D system and a conventional Josephson junction (JJ).  In a classic 
experiment, Eck, Scalapino and Taylor\cite{eck} observed resonances at finite 
voltage in the dc tunnel current of a JJ in the presence of a parallel field 
$B_{||}$.  These resonances were successfully interpreted as resulting from 
excitation of the electromagnetic modes (i.e. the Swihart modes) of the
junction  by the ac Josephson current.  At non-zero $B_{||}$ the temporal and
spatial  oscillations of the Josephson current can resonantly excite the
junction modes.  This leads to features in the dc tunneling characteristics
which trace out the  linear dispersion of those modes as the parallel field is
varied. In the present  case the analogous mode is the pseudospin Goldstone mode
and the corresponding  excitation is produced by the periodic tunneling currents
which result from the  ferromagnetic order within the bilayer 2DES at finite
$B_{||}$\cite{girvin}. 

In summary, magneto-tunneling spectroscopy experiments on double layer 2D 
electron systems in the $\nu_T=1$ QHE state reveal a collective mode in the
system which disperses linearly with wavevector at low energy.  The measured 
velocity of this mode is in reasonable agreement with theoretical estimates for 
the Goldstone mode of the broken symmetry ground state.  The question of whether
the system supports a Josephson effect remains open.

It is a pleasure to acknowledge fruitful discussions with Steve Girvin, Allan 
MacDonald, Ady Stern and Misha Fogler.  This work was supported by the NSF
under  Grant DMR0070890 and the DOE under Grant DE-FG03-99ER45766.  One of us
(I.B.S.)  acknowledges support from the Department of Defense.


\begin{references}
\bibitem{true spins} The true spins are assumed to be fully polarized by the 
Zeeman energy.
\end{references}
\end{document}